# A Review of Macroscopic Motion in Thermodynamic Equilibrium

Juan R. Sanmartin

(Universidad Politécnica de Madrid)

Abstract

A principle on the macroscopic motion of systems in thermodynamic equilibrium, rarely discussed in texts, is reviewed: Very small but still macroscopic parts of a fully isolated system in thermal equilibrium move as if points of a rigid body, macroscopic energy being dissipated to increase internal energy, and increase entropy along. It appears particularly important in Space physics, when dissipation involves long-range fields at Electromagnetism and Gravitation, rather than short-range contact forces. It is shown how new physics, Special Relativity as regards Electromagnetism, first Newtonian theory then General Relativity as regards Gravitation, determine different dissipative processes involved in the approach to that equilibrium.

## I Introduction

I will start this *Review* by recalling that *rigid-body* motion is a definite property of the thermodynamic equilibrium of macroscopic systems, a consequence of maximum entropy requiring maximum internal energy and minimum macroscopic energy $E_M$, this being a characteristic of rigid-body motion: Very small but still macroscopic parts of a fully isolated systems in thermal equilibrium move as if points of a rigid body [1a, b].

Surprisingly, thorough *introductory treaties*, such as the 3-volume *Feynman Lectures* or the 5-volume *Berkeley Course*, and advanced *treaties* such as the 6-volume *Sommerfeld Lectures*, fully ignore the above property of thermodynamic equilibrium. This is not the case of the 10-volume *Landau-Lifshitz Course*, which explicitly brings it forth in its 5[th] volume, though in a limited way [2].

Purpose of this Review is showing how new physics, emerging at different historical points and particularly applying in *Space Physics*, keeps agreeing with the above macroscopic equilibrium condition, different, new dissipative processes being involved in the approach to this equilibrium: Dry friction, through short-range force, is recalled in Sec. II; Newtonian gravitation, through tidal forces in Solar planet/moons systems, is reviewed in Sec. III; Special Relativity, through Electromagnetism and *Tether* applications, in Sec. IV;

and General Relativity, through *Lense-Thirring* frame dragging and *Black Holes* merging in Sec. V. Conclusions are given in Sec.VI.

II Surface-contact friction

The simplest case of contact forces, dry friction of solids on solids, might have been the oldest one allowing direct observation of a dissipation mechanism. Macroscopic energy is then just kinetic energy, the case considered in Ref. 2, with $E_M$ minimum at given momentum and angular momentum. Any friction law that keeps acting as long there is relative sliding involves thermodynamic finality.

Time to reach equilibrium may affect the dynamics, however; a dropped ball may bounce a number of times before reaching relative rest, at the floor. Actually, dry friction is thermodynamically peculiar. A solid may support shear stresses (and strains) in metastable equilibria near condition of actual equilibrium, for times that may be considered infinite; entropy can then depend on strain or shear variables instead of just volume or pressure [1a], [3].

As shown in Sec.III, in systems involving gravitation *rigid body* motion may correspond to either maximum or minimum entropy, one of the thermodynamic paradoxes of gravitation, along with *Jeans* instability and *negative* heat capacity of systems. Also, a system may be trapped in a metastable entropy maximum for very long times. Further, systems may also exhibit certain entropy *runaway*.

One can also find dry-friction equilibria with two energy extrema, as in a case involving conserved angular momentum and just gravity [1a], but no paradox is involved: A particle of mass $m$ rests inside a bowl of symmetric shape $z_b(r) \propto r^2$ and moment of inertia $I$, which turns freely around its vertical axis at given angular momentum $L_0$. There are energy extrema at values

$$v_r = 0, \qquad L_0^2 r = \left(I + mr^2\right)^2 g \frac{dz_b}{dr}, \qquad v_\theta = \frac{L_0 r}{I + mr^2}$$

and at the bowl bottom. They are both minima, however; further, they exist at different parameter ranges, the first and second minima above corresponding to $(L_0/I)^2$ greater and smaller than $gd^2z_b/dr^2$ (*a constant*), respectively.

III Tidal forces in Newtonian Gravitation

The Newton law of gravitation was used in the late XVIII century to study the effects of tidal forces on the orbital evolution of planet/satellite systems in the Solar system, including the case of multiple satellites as presented by Jupiter, through extensive, deep mathematical analyses by P. S. Laplace [4]. The long-range gravitational forces, contributing to macroscopic energy $E_M$, involve large enough bodies that introduce tidal dissipation arising from body or ocean tides. Tidal forces would not occur in case of gravitation among just point masses.

Such forces cover the quite broad range of celestial-body motions, with resonances that were studied when *Thermodynamics* was still coming into being; language then used makes typically no reference to their thermodynamic character. Synchronization of rotation / revolution periods for 2/3 body-systems, called orbital resonances, may require extremely long times and then manifests itself through partial equilibria. Particular equilibria may show sustained chaotic dynamics.

Considering a planet/satellite system, orbital motion and planet and satellite spins contribute to both energy $E_M$ and angular momentum $H_0$, reading, in the simplest description,

$$E_M = \frac{1}{2} I_p \omega_p^2 - \frac{1}{2} \frac{\mu p M_s}{a} + \frac{1}{2} I_s \omega_s^2, \qquad \mu p \equiv GM p$$

$$H_0 = I_p \omega_p + \frac{\mu p M_s}{a\hat{\Omega}} \sqrt{1 - e^2} + I_s \omega_s = constant$$

where $I_p$, $I_s$ are the largest principal moments of inertia, with both spin axes assumed perpendicular to the orbital plane; $\hat{\Omega}$ is the mean orbital angular velocity; and $a$ is the major semi-axis. Use of the $H_0$ relation and *Kepler*'s 3$^{rd}$ law,

$$a^3 \hat{\Omega}^2 = \mu p + GM s,$$

allow considering $E_M$ is function of just three convenient variables, say, $\omega_p$, $e$, and $a$.

Conditions for extrema are readily found to read

$$e = 0 \qquad \omega_s = \hat{\Omega} = \omega_p$$

Complete equilibrium thus corresponds to circular orbit and full synchronism, a condition reached by *Pluto* and its (large) moon *Charon*, which were recently found to fully move as a single rigid body, with all three periods being 6.387 days. It does correspond to minimum $E_M$ and thus to a maximum of entropy, in a stable equilibrium. Pluto-Charon is the only known system that has reached this full synchronism, the stable end state of tidal evolution, with the Pluto and Charon equatorial planes both coinciding with the plane of their mutual orbit.

Pluto and its moon Charon have comparable sizes and masses. Usually, however, small ratios $I_s/I_p$, $M_s/M_p$, and $I_s/M_s a^2$ allow a value $d\omega_s/dt$ much larger than $d\omega_p/dt$ and $d\hat{\Omega}/dt$, and the satellite spin to first approach synchronism with a 'frozen' orbital motion, while the eccentricity is reduced [5]. Consider a light satellite (including the case of an artificial, extremely light satellite) and assume that values $e = 0$ and $\omega_s = \hat{\Omega}$ ($\equiv \Omega$), have already been reached in a first stage (*spin-orbit* resonance), drop the last term in the $E_M$, $H_0$, and Kepler equations, and set $H_0 > 0$ with no loss of generality, a retrograde, $\Omega < 0$ satellite then requiring positive $\omega_p$.

The equation for an $E_M$ extremum condition can now be rearranged to read

$$\frac{\partial E_M}{\partial a} = \frac{M_s}{2}\sqrt{\frac{\mu_p}{a}}\left(|\Omega| \mp \omega_p\right) = 0,$$

the lower sign corresponding to negative $\Omega$. There is no entropy extremum in that case, when the parenthesis above is positive along with $\omega_p$. $E_M$ is then found to decrease with decreasing distance $a$ through the term $-\mu_p M_S/2a$, and entropy to increase monotonically. This is the case of moon Triton, which is in retrograde motion relative to Neptune; it should finally fall into its planet.

For the prograde case, the extremum condition above, showing that any entropy maximum will correspond to synchronous periods of planetary rotation and satellite revolution, can be rewritten as

$$\frac{\partial E_M}{\partial a} = \frac{M_s \mu_p}{2 I_p \sqrt{a}}\left(\frac{I_p}{a^{3/2}} + M_s \sqrt{a} - \frac{H_0}{\sqrt{\mu_p}}\right) = 0.$$

This can be seen as determining the $a$ value(s) at energy extrema for given $H_0$ positive, but can also be seen as determining $H_0$ as function of (real) $a$. This shows that there exist positive $\Omega$ (and $H_0$) cases that present no extremum condition.

Indeed, $H_0(a)$, so defined for $\Omega > 0$, does have a positive minimum as function of $a$,

$$\min H_0 \equiv H^* = 4\left(I_p \mu_p^2 M_S^3/27\right)^{1/4}$$

at $a = \sqrt{3I_p/M_S}$. There is no extremum for the $\Omega > 0$, $H_0 < H^*$ case, which is then similar to the case $\Omega < 0$, whereas for $\Omega > 0$, $H_0 > H^*$, $E_M(a)$ presents two extrema, a minimum, and a maximum *closer* to the planet, where entropy is minimum, which makes it unstable. Note that artificial Earth satellites (in *prograde* orbit) do have $H_0 \gg H^*$ and thus an $a(max)$ equilibrium, the unstable geostationary orbit; this is 'why' such satellites re-enter, escaping from the far-away *rigid-body* co-rotation with Earth, at geostationary, 'using' short range forces (air drag). For artificial satellites $a(min)$ might be larger than the estimated Universe size.

Also, the equilibrium at $a(min)$ would be formally metastable rather than stable because the $E_M$-minimum is relative; to the left of the maximum, entropy increases indefinitely with decreasing $a$, as in entropy runaway. The satellite, if between $a(max)$ and $a(min)$, recedes away from the planet, taking energy from a decrease in the planet spin. For planets with several moons, however, a receding-moon process may lead to its being caught into a *Laplace resonance*, as found in the complex moon system of Jupiter.

This is not the case of the Earth single moon. As the Moon would move away from $a(max) \sim$ 14,500 km (if ignoring the prevalent theory of the Moon origin through the impact of a *Mars*-size object with Earth), towards $a(min) \sim$ 550,000 km, both (equal) moon spin $\omega_s$ and orbital angular velocity $\Omega$ would slow down, as determined by Kepler's law, $a^3\Omega^2 = \mu_p$, with the Earth spin slowing down somewhat too, as following from conservation of angular momentum, $H_0 = I_p\omega_p + M_s\sqrt{GM_p a}$. Forces from tidal bulges are responsible for that evolution. As the satellite rises a tide on its planet, they exchange angular momentum. This changes the satellite orbit and the planet spin; the Moon gains orbital energy at the expense of the Earth rotational energy.

In the assumed remote past, when the Moon was closest, both month and day would have been as short as 4 – 5 hours. At $a(min)$, where a Pluto/Charon condition would hold, all three periods (Earth and Moon rotations, and orbital revolution) would be equal at about 47 days. Actually, its high, 0.055, eccentricity shows that the Moon has not gone fully through its first stage, and is not quite in the synchronous resonance. Because the distance from Moon to Earth varies along the orbit, its angular velocity is not quite constant; only its average is equal to its spin, $\omega_s = \hat{\Omega}$.

IV Electromagnetism and Special Relativity

The case of tethers involves dissipation arising from *Relativity* physics. Consider two macroscopic systems *S* and *S'* approaching in free motion in certain space-time domain [1b], *S* moving with velocity $\mathbf{V_{rel}}$, constant and uniform, relative to *S'*; both systems could stand for inertial reference frames. The velocity $\mathbf{V_{rel}}$ may now be looked at in two respects. First, $\mathbf{V_{rel}}$ by itself might give rise to reduction of mechanical energy by any dissipative mechanism available to make *S* and *S'* move as a single rigid-body. Secondly, a Lorentz transformation of fields may explicitly introduce $\mathbf{V_{rel}}$ in the analysis of *S* motion, say, actually introducing such a mechanism.

Let system *S'* be a magnetized, highly conductive plasma, and *S* a conductive wire travelling through *S'*, and let $V_{rel}$ be much smaller than the light speed *c*. The electric field in the plasma will then vanish in its own frame, $E' = 0$, whereas in the *S* frame will approximately read

$$\vec{E} = \vec{E}_m \equiv \vec{V}_{rel} \times \vec{B}_0,$$

where $E_m$ is often called *motional* field and $B_0$ is the ambient magnetic field, which keeps about equal in both frames.

This motional field, which may drive a current in the wire, underlies the physics involved in the applications of ED (*Electrodynamic*) space tethers. In the simplest case of that current coming out weak, the wire would be near equipotential in its own frame. The wire will present anodic and cathodic segments that need to allow overall balance of electron collection and ejection. Note that the Lorentz force on the wire current makes the field $B_0$ enter that drag twice, making its thermodynamic, unidirectional character manifest.

The story of tethers might start in 1960, when *Beard* and *Johnson* discussed the magnetic drag on a satellite as a conductor moving across the geomagnetic field lines [6], in XIX *century* physics language, kept later in most tether work; naturally, M. Faraday could not know about plasmas and Langmuir probes. Beard and Johnson estimated that a voltage of order 200 V/km could be induced, and the magnetic drag on the resulting current would exceed the air drag at altitudes above 1200 km for satellite sizes over 50 m; they concluded that magnetic drag effects could be ignored.

A new analysis of the problem of a conductor moving across a magnetic field was carried out by *Drell et al* in 1965 [7]. They straightforwardly considered current as arising from the electric field as seen by a co-moving observer. They suggested that the main impedance in the current circuit, closed through the ambient ionosphere, was radiation impedance of *Alfven* waves. They called drag from such emission Alfven propulsion; it was not propulsion introduced by H. Alfven. Actually, impedance for steady emission by a tether depends on the way it exchanges current with the ambient plasma. Used models of tether current flowing from $y = -\infty$ to $y = +\infty$ if lying along some axis $y$, say, did underestimate impedance by a factor of order $1/n^2$ ($n$ being the very high refraction index), as against taking into account that current vanishes at a finite tether ends.

*Drell et al* implied that current did easily flow in and out of wire and ambient plasma. In work presented in early 1966, *R. D. Moore* thought otherwise and discussed ambient plasma-wire contact impedances in the current circuit. He argued the need for using plasma-contactor devices such as used in electric propulsion to neutralize ion engines exhaust [8]. He also realized that a kilometres long wire might be needed to attain a large *motional* electromotive force and suggested that a gravity-gradient force could keep the wire vertically stable in Low Earth Orbit. Moore was somehow setting the foundations of the new technology of ED tethers for use in LEO.

*Moore* went further in mid-1966 and took the *motional* field concept to the Solar Wind plasma, although, with no gravity-gradient force available there, he tried to escape the need for a long wire [8]. Alfven first considered a propulsion problem in 1972 [9a], while visiting at the University of California at San Diego, and again later that year, with U. *Fahleson* at the Royal Institute of Technology in Stockholm [9b]; in Note added in proof in this last paper, Alfven apologized for their ignorance of Moore's work 6 years earlier.

There is, anyway, a basic difficulty in using the *motional* Lorentz drag in the Solar Wind, that being its extremely low values of both ambient plasma density and magnetic field. This has given rise to the alternative electric solar-sail concept, using an array of wires biased by solar power to deflect Solar Wind ions, hence generating drag/thrust through Coulomb, rather than Lorentz, forces [10], to overcome just the weak ambient magnetic-field condition. Note that this scheme does require use of an independent power source, and lacks the thermodynamic, ready-to-use character of Lorentz drag.

Tethers first advanced on the basis of an insulated wire and a big spherical ($R \sim 1$ m) conductor as passive anodic collector. This was supposed to apply in ambient plasmas with *Debye* length of order of 1 centimeter... In 1993, the *bare tether* concept [11], leaving the <u>thin</u> wire uninsulated to allow it to directly collect electrons over segments of order of kilometres, with no need for anodic-end contactor, completed the concept. The *bare-tether* collection scheme of Ref. 11, was acknowledged as explaining a phenomenon on board the International Space Station in 2001 [12]. Results from a multiple-scales asymptotic analysis of tethers operating at each tether point as Langmuir probes [13] was verified in a detailed numerical code in a University of Michigan PhD Thesis by Eric Choinière [14].

De-orbiting satellites at end of life, so as to prevent generation of new space debris, is a paradigmatic mission exhibiting the thermodynamic character of electrodynamic tethers, playing the same passive role of air drag. Space debris remain a constant menace to operative satellites in Low Earth Orbit. Results from a basic analysis on the *tape* geometry of tethers as de-orbit systems to be used just at end of mission, follow work on overall development of tether systems, under a Project, *BETs*, supported by the *European Commission* from November 2010 through February 2014 [15], [16].

Any de-orbiting system faces two basic requirements: it must be light when compared to its satellite, and operate fast to avoid its accidental, catastrophic collision with another large orbiting object, resulting in a myriad of debris pieces. A tether system also faces three additional issues: it might be somewhat ineffective at high inclination orbits for which $E_m$ could prove too low; its long, thin geometry makes it prone to cuts by abundant tiny debris, leading to a failed operation; and its geometry (long) make it also prone to cut by a big debris.

Use of thin tapes [13] instead of round wires was a further improvement on the bare tether concept. Because a tape has larger cross-section perimeter than a (*fair-comparison*) round wire of equal mass and length, it is a more efficient anodic collector and de-orbits faster. Also, it is less prone to small debris cuts because present debris flux decreases fast with larger debris size, as required by a tape cut. Overall, recent results showed that tape tethers have much greater survival probability than round tethers of equal length and mass [17].

Tether geometry has thus a relevant impact on system performance, and tape tethers are advantageous in this respect.

Given a mission, i.e. initial orbital parameters and satellite mass, one might choose tether length, width, and thickness to optimize figures of merit. The opposite requirements of both a light tether and survivability against debris lead to a design scheme based on the product of probability of a cut by debris and tether-to-satellite mass ratio $m_t/M_S$. A complementary figure of merit is a product involving de-orbit time $t_f$ and ratio $m_t/M_S$. Proper tether design with very small mass-ratios leads to very short de-orbit times at both low and mid-inclination orbits. Moving to polar orbits, mass ratios just twice as large allow for $t_f$ (still reasonably) 4 times as long [16].

The thermodynamic character of tether operation is also manifest at *capture / apojove lowering* missions to the Giant Outer planets, for convenient, full exploration of their moons. A recent application to Jupiter's moon *Europa*, leading to tens-of-kilometre long tapes, which Lorentz drag might need because of low ambient plasma density, first resulted in electrons (intended to be collected) actually reaching the tape with energy so high that *range* (penetration length) $\delta_e$ exceeded tape thickness [18].

Final design set *thickness* = $\delta_e(\varepsilon_{max})$, with $\varepsilon_{max}$ the maximum energy of electrons reaching the anodic tether-end during perijove passes, throughout operation just hundreds of kilometres above the planet. The result is thin and short tethers, that capture spacecraft several times as heavy, just 200-300 kg, say, allowing for apparently easy missions. Strong heating of the tape might be solved by surface treatment of aluminum to increase its thermal emissivity while keeping it highly conductive.

Need for reducing costs of space missions has been long a pressing one. The Direction of NASA's Planetary Science Division recently proposed designing and flying robotic space probes to so called *Ice giants*, Uranus and Neptune, with common space platforms (two copies). It insisted on looking at scaled back concepts to be developed <u>at less cost</u>, and identifying potential concepts across a spectrum of price points.

Missions to all 4 Giant planets, Jupiter, Saturn, Uranus, and Neptune, face common issues. They are far from the Sun, and present deep gravitational wells, far from the Earth, setting both *power* and *propulsion* issues. Solar power might be hardly effective, and *poor* available power, making electric propulsion unfeasible, makes for a pressing propulsion issue, requiring huge wet mass for rockets, if not just a flyby mission but to operate through the planet gravitational well.

Fortunately, all *Ice* and *Gas* Giants have magnetic field and ambient-plasma electrons, allowing for common mission concepts (as well as *Rings* and *Radiation Belts*, as common secondary issues...). Electrodynamic tethers can provide propulsion using no propellant for both planetary capture and operation down the gravitational well. They could also generate power along, for use in those operations, or in storing to invert the tether current in operating up the well.

V  General Relativity and Black Holes

The fate of a star is very much dependent on its mass. The *Jeans* condition for gravitational collapse requires, in a gross way, that gravitational energy of a cloud of ideal gas be greater than its thermal energy. For mass $M$, radius $R$, and number $N$ of particles of average mass $\overline{m}$, an approximate condition reads

$$GM^2/R > Nk_BT \qquad \text{i.e.} \qquad \rho > \left(k_BT/G\overline{m}\right)^3/M^2.$$

Since minimum density decreases rapidly with cloud mass, formation of stars could occur in stages. First, a cloud having, say, thousands of times a Sun mass and density as above, does collapse. Once density becomes high enough, much less massive parts may each collapse into a *protostar*, leading to a *star cluster*.

There are lower and upper bounds to mass of possible stars. At some point, contraction proceeds in hydrostatic equilibrium, with pressure balancing gravitation under the *virial* theorem. At given mass $M$, the star heats as density increases. Heating stops, however, at density so high that quantum mechanics sets in, with (degenerate) electron wave-functions starting to overlap. That density reads

$$\rho \approx \overline{m}(m_e k_B T)^{3/2}/(2\pi\hbar)^3.$$

When equated to the minimum *Jeans* density above, it yields a maximum temperature

$$k_B T_{max} \sim \frac{G^2 \overline{m}^{8/3} m_e}{4\pi^2 \hbar^2} \times M^{4/3}.$$

For a protostar to become a star, that temperature must be above a minimum temperature $T_{ign}$ triggering thermonuclear reactions that fuse protons into helium nuclei, with $\overline{m} = 0.5$ $m_H$ for ionized hydrogen; $k_B T_{ign}$ is some small fraction of the electron rest mass, $m_e c^2 \approx$ 0.5 MeV, about $1.5 - 2 \times 10^6$ K. This determines a minimum mass for stardom

$$M_{min} \approx 0.05 \, (\hbar c / G)^{3/2} / m_H^2 \sim 0.1 \, M_{Sun.}$$

A protostar with $M < M_{min}$ (*brown dwarf*) does not make it to a star; it keeps too cold, with degenerate-electron pressure balancing gravitation.

At the opposite end, at large enough *M*, radiation (photon) pressure dominates gas pressure, their ratio reading $p_{rad} / p_{gas} \propto M^2$. The virial theorem for ultra-relativistic particles (case of photons moving at speed *c*) shows that as *M* increases the *binding* energy rapidly decreases. Stars with *M* above, say, 50 $M_{Sun}$ have binding so weak that they are nearly unstable and rarely found.

Stars spend most of their life burning hydrogen. When hydrogen in the central nuclear-burning region (the core) is depleted, that core contracts at increasing density and temperature. The star may then switch to a new stage, burning helium. Depending on its initial mass, the star may go through new, successive stages, each time burning heavier elements though not going further than iron, which has the highest binding energy per nucleon. Each new burning stage proceeds faster and faster and *cooks* heavier elements.

Stars with about 0.5 - 5 solar masses evolve into *Red giants* when core hydrogen is exhausted. The core, contracting under its gravity, heats an outer shell where fusion of hydrogen to helium then starts. The star luminosity increases by a factor of $10^3$-$10^4$ and the outer layers greatly expand. Generated energy spreads over a much larger surface area, resulting in lower surface temperature and an emission shift to red-light… In stars massive enough, helium fusion is ignited at the core; when that helium is exhausted, an analogous process leads to an *asymptotic red-giant* phase.

The Sun will become a red giant engulfing the solar system's inner planets up to Earth. Most of the atmosphere would get lost in space; ultimately, Earth would be a desiccated, dead planet with a surface of molten rock. For not too high mass and as part of a red-giant phase, the star will eject a cold outer layer forming a *planetary nebula* with its core ultimately exposed as a *White Dwarf*. There is a maximum value to the mass of a *white dwarf*, as given by the *Chandrasekhar* value,

$$M_{Ch} \propto (\hbar c/G)^{3/2} / m_H^2, \quad \text{about} \quad 1.4 M_{Sun}$$

In so called *Neutron stars*, densities are much larger than in white dwarfs, with gravitational forces balanced by the pressure of degenerate neutrons. Neutron stars already involve General Relativity, with *Energy* $\sim Mc^2$ Due to neutron-interaction and General Relativity effects, $M_{max}$(*neutron star*) may be about 3 $M_{Sun}$. In the route to a white dwarf or a neutron star condition, a star sheds entropy in addition to (binding) energy and mass, degenerate-particle pressures corresponding to both negligible temperature and entropy.

Apparently, no pressure can prevent a collapsing mass greater than about $3M_{Sun}$ from reaching its *Schwarzschild* radius, $R_S(M) \equiv 2GM/c^2$. Purely dynamical results then show singular thermodynamic behaviour. The simplest (*Schwarzschild*) *black hole*, having neither angular momentum nor electric charge, is characterized by just its mass *M*.

Once it drops below $R_S$, the star would collapse all the way to a central singularity at infinite density and space-time curvature. An unlimited runaway of entropy would occur at collapse. The surface at $R_s$ is an event *horizon*: no particle, no light ray, indeed no signal could leave its inside; they can get in from outside. Thus, entropy generated during collapse cannot be shed. This is a fundamental difference with *white dwarf* / *neutron* stars.

A bound to the central singularity would arise from quantum effects on space-time curvature at distances of the order of the Planck length, $\sqrt{G\hbar/c^3}$ ~ $10^{-35}$ m. Entropy should then saturate at extremely high values, quantum effects being essential in yielding a definite value for entropy. For external observers a black hole is characterized by a single, mass, parameter. For instance, the external fields of any initial non-sphericity are wiped out during collapse.

One can just write $S_{bh}(M)$ or $S_{bh}(R_s)$, and determine it from the very definition of entropy as additive quantity, and thus presumed extensive, even if energy or mass is not. For observations conducted entirely outside the *horizon*, the black hole is located there, and so its extension is its area $4\pi R_s^2$. Entropy, being thus proportional to $M_s^2$ too, had been presumed non-extensive.

Although the horizon is a globally defined construct of General Relativity, with no special significance for local observers, quantum effects, proved essential for the existence of entropy, could change this. There are suggestions that entropy lies in a sheath around the horizon, its thickness, about a *Planck* length, collapsing as $\hbar \to 0$, hence, keeping entropy off the black-hole exterior in the classical limit.

From $S_{bh} \propto R_s^2 \propto M^2 \propto E^2$ and the definition of temperature we have

$$1/T \equiv \partial S_{bh}/\partial E \propto E \quad \Rightarrow \quad TS_{bh} = \tfrac{1}{2} E,$$

an increase in the energy of a black hole decreasing its temperature. This is in agreement with *negative* heat capacities as a generic feature of gravitationally bound systems. Since only universal constants, in addition to $R_s$, or $M$, or $E$, might enter the dimensionless expression for entropy, we could write [19],

$$\frac{S_{bh}}{k_B} = \alpha \frac{R_s^2}{G\hbar/c^3} \equiv \alpha \frac{4GM^2}{\hbar c}$$

with $\alpha$ some dimensionless factor (found to be $\pi$). Black-hole entropy is generally enormous. A black hole with the Sun's mass ($1.99 \times 10^{30}$ kg) would have $S_{bh} \approx 1.45 \times 10^{54}$ J / K. This compares with the present Sun entropy, which is about $10^{35}$ J / K.

A number of simple thermodynamic results follow. From $T = E/2S_{bh}$ one readily finds $k_B T \equiv \hbar c/4\pi R_s$, which can be read as $k_B T = \hbar \times$ *a frequency*. Radiation thermodynamics suggests that a black hole radiates as a black body at the above temperature. This rests on a law $S_{bh} \propto R_s^n$, <u>$n = 2$</u>, with the previous $\pi$ factor then determined. The dependence $S_{bh} \propto 1/T^2$ ($S_{bh} \to \infty$ as $T \to 0$), shows it opposite Planck's 3[rd] law of thermodynamics, $S_{bh} \to 0$ as $T \to 0$.

Recently, there was a first direct evidence of a binary black-hole *merger* [20], through the first direct detection of a transient gravitational signal (GW150914), as emitted in the process. As against Newtonian gravitation, which could give rise to no gravitational waves (Alfven wave radiation earlier discussed for tethers was entirely electromagnetic), General Relativity does naturally allow for waves, while not allowing for elliptical orbits, two objects in relative orbit slowly spiraling in, as energy is radiated away. Black hole merging appears as a General Relativity mechanism to achieve rigid-body motion.

The energy and entropy changes in the merger can be directly exhibited. An extremely thorough measurements evaluation led to mean-value masses of the merging black holes estimated as $M_1 = 36\ M_{Sun}$ and $M_2 = 29\ M_{Sun}$ and the final black hole mass $M_f$ as $62\ M_{Sun}$, a total energy $3M_{Sun}c^2$ having been radiated away. For Schwarzschild black holes there would then be a change in entropy

$$\Delta S_{bh} \propto \Delta M^2 = [62^2 - (36^2 + 29^2)] \, M_{Sun}^2 = 1707 \, M_{Sun}^2,$$

a positive value, as against the mass loss in the merger, because of the mass *quadratic* dependence of entropy.

Actually, *Kerr* black holes, as characterized by both mass and angular momentum $J$, always limited to the range $J < Mc \times \frac{1}{2} R_S$, appear to be involved. Kerr and Schwarzschild black hole entropies are similar, with just the change to

$$\frac{S_{bh}}{k_B} = \frac{2\pi GM^2}{\hbar c} \times \left[1 + \sqrt{1 - (2J/McR_S)^2}\right],$$

recovering the Schwarzschild case for $J = 0$ (and singularly excluding the extremal case $J = Mc \times \frac{1}{2} R_S$). The normalized angular momentum $2J / McR_S$ was estimated as 0.67 for $M_f$ and just less than 0.7 for $M_1$ with no estimate for $M_2$. Since dealing with the merging black holes as of Schwarzschild type will not underestimate their entropy, writing

$$\Delta S_{bh} \propto 62^2 \times \frac{1}{2}\left[1 + \sqrt{1 - 0.67^2}\right] - 36^2 \times \frac{1}{2}[1+1] - 29^2 \times \frac{1}{2}[1+1] = 1207 > 0$$

does not overestimate the entropy increase.

We note, in concluding, that a Kerr black hole exhibits the *Lense-Thirring* (*dragging of inertial frames*) effect of General Relativity, that makes rigid-body motion manifest [21], and also, supermassive black holes may carry tidal disruptions of Newtonian Gravitation to the extreme [22].

VI Conclusions

We reviewed a rarely discussed principle on the macroscopic motion of systems in thermodynamic equilibrium, very small but still macroscopic parts of fully isolated systems in thermal equilibrium moving as if points of a rigid body, macroscopic energy of motion being dissipated to increase internal energy, and increase entropy along. We have illustrated the new ways fundamental physics at long-range fields (Special Relativity for Electromagnetism, first Newtonian theory then General Relativity for Gravitation), of particular interest in Space physics, introduced ways to dissipate energy different from processes at short-range contact forces. Newtonian gravitation involves dissipation at tides arising from the distance-dependent forces throughout macroscopic bodies; General

Relativity involves effects very weak outside Astrophysics, *Lense Thirring* frame-dragging by rotating bodies, merging of bodies in relative orbiting by inspiral due to gravitational-wave radiation, which is fully absent in Newtonian gravitation. Electromagnetism involves the two-steps effect of magnetic fields that induce electric fields in inertial frames at relative motion, then exert Lorentz-drag on ensuing currents.